\documentclass{article}
\usepackage[utf8]{inputenc}
\usepackage{authblk}
\usepackage{setspace}
\usepackage[margin=1.25in]{geometry}
\usepackage{graphicx}
\usepackage{fancyhdr}
\usepackage{longtable}
\usepackage{parskip}
\usepackage{subcaption}
\usepackage{CJK}
\usepackage{bm}
\usepackage{color}
\usepackage{amsfonts}  
\usepackage{amsmath}   
\usepackage{amsthm,amscd,amsxtra,amsfonts,amsmath,amssymb}
\usepackage{multirow}

\title{Molecular geometric deep learning}

\author[1,2]{Cong Shen}
\author[1*]{Jiawei Luo}
\author[2*]{Kelin Xia}

\affil[1]{College of Computer Science and Electronic Engineering, Hunan University, Changsha 410000, China}
\affil[2]{School of Physical and Mathematical Sciences, Nanyang Technological University, 637371, Singapore}
\affil[*]{Corresponding author. Email: luojiawei@hnu.edu.cn, xiakelin@ntu.edu.sg}

\date{}

\begin{document}

\maketitle

\begin{abstract}
Geometric deep learning (GDL) has demonstrated huge power and enormous potential in molecular data analysis. However, a great challenge still remains for highly efficient molecular representations. Currently, covalent-bond-based molecular graphs are the de facto standard for representing molecular topology at the atomic level. Here we demonstrate, for the first time, that molecular graphs constructed only from non-covalent bonds can achieve similar or even better results than covalent-bond-based models in molecular property prediction. This demonstrates the great potential of novel molecular representations beyond the de facto standard of covalent-bond-based molecular graphs. Based on the finding, we propose molecular geometric deep learning (Mol-GDL). The essential idea is to incorporate a more general molecular representation into GDL models. In our Mol-GDL, molecular topology is modeled as a series of molecular graphs, each focusing on a different scale of atomic interactions. In this way, both covalent interactions and non-covalent interactions are incorporated into the molecular representation on an equal footing. We systematically test Mol-GDL on fourteen commonly-used benchmark datasets. The results show that our Mol-GDL can achieve a better performance than state-of-the-art (SOTA) methods. Source code and data are available at https://github.com/CS-BIO/Mol-GDL.
\end{abstract}


\section{Introduction}
Artificial intelligence (AI) based models have demonstrated huge power and enormous potential for the prediction of various molecular properties. In particular, AI-based drug design has achieved great success in various steps in virtual screening and has the potential to revolutionize the drug industry \cite{zhang2017machine,chen2018rise,mak2019artificial,chan2019advancing}. However, even with the great progress,  to design efficient molecular representations and featurization remains a great challenge. In general, all AI-based molecular models can be classified into two types, i.e., molecular descriptor-based machine learning models and end-to-end deep learning models.

The first type is to use handcrafted molecular descriptors or fingerprints as input features for machine learning models. The generation of the handcrafted descriptors is known as featurization or feature engineering. Other than physical, chemical, and biological properties, such as atomic partial charge, hydrophobicity, electronic properties, steric properties, etc, the majority of the molecular features are obtained from molecular structural properties. In fact, more than 5000 molecular structural descriptors have been developed, and can be generalized as one-dimensional (1D), two-dimensional (2D), three-dimensional (3D), and four-dimensional (4D) features \cite{puzyn2010recent,lo2018machine}. The 1D molecular descriptors include atom counts, bond counts, molecular weight, fragment counts, functional group counts, and other summarized general properties. The 2D molecular descriptors include topological indices, graph properties, combinatorial properties, molecular profiles, autocorrelation coefficients, etc. The 3D molecular descriptors include molecular surface properties, volume properties, autocorrelation descriptors, substituent constants, quantum chemical descriptors, etc. A related higher computational cost is usually required for the generation of 3D molecular descriptors. The 4D chemical descriptors characterize configuration changes in a dynamic process. These structural descriptors are widely used in quantitative structure-activity relationships (QSAR), quantitative structure-property relationships (QSPR), and machine learning models. Recently, deep learning models, including autoencoder, convolutional neural network (CNN), and graph neural network (GNN), have also been used in molecular fingerprint generation \cite{merkwirth2005automatic,duvenaud2015convolutional,coley2017convolutional,xu2017deep,winter2019learning}. For all machine learning models, their performance is directly related to the efficiency of these molecular descriptors.

The second type is end-to-end geometric deep learning (GDL) models. In these GDLs, molecules are represented as molecular graphs, density functions, or molecular surfaces, and various deep learning models, such as (3D) convolutional neural networks (CNNs), graph neural networks (GNNs), recurrent neural networks (RNNs), etc, can be used to automatically learn the molecular properties \cite{wieder2020compact,yu2022molecular,atz2021geometric,li2022geomgcl,wang2022molecular}. Among these different molecular representations, molecular graphs are the most popular one. In particular, covalent-bond-based molecular graphs are the de facto standard for representing molecular topology at atomic level. Based on them, various GDL models have been proposed, including graph recurrent neural networks (GraphRNN) \cite{pmlr-v80-you18a}, graph convolutional networks (GCN) \cite{welling2016semi}, graph autoencoders \cite{kipf2016variational}, graph transformers \cite{yun2019graph}, and others. These GDLs have been widely used in molecular data analysis, in particular drug design \cite{kotsias2020direct, wang2021multi, vamathevan2019applications}. Recently, non-covalent-interaction-based molecular descriptors have achieved great performance in the prediction of binding affinities of protein-ligand and protein-protein \cite{wang2020topology, meng2021persistent}, demonstrating potential new molecular graph representations beyond the de facto standard of covalent-bond-based graphs. In fact, the incorporation of geometric information, such as bond angles, periodicity, symmetry, rotation-translation invariance, equivalence, etc, into GDL models, can help to significantly improve the learning performance \cite{schutt2018schnet,GONG2021110332,kim2018deep,batatia2022mace}, especially for the machine learned based force field models. Efficient representations beyond covalent-bond-based molecular graphs provide a great promise for a better representation of molecular geometric information.


Here we show that molecular representations with only non-covalent interactions can achieve similar or even better results than the de facto standard of covalent-bond-based models in molecular property prediction, for the first time. More specifically, we systematically compare the performance of GDL models using two types of molecular representations, i.e., covalent interaction graphs and non-covalent interaction graphs, in several most-commonly-used benchmark datasets, including BACE, ClinTox, SIDER, Tox21, HIV, ESOL, etc. It has been found that GDL models using only non-covalent interactions have comparable or even superior performance than the de facto standard models. Further, we propose molecular geometric deep learning (Mol-GDL) to incorporate a more general molecular representation into GDLs. In our Mol-GDL, molecular topology is modeled as a series of molecular graphs, each focusing on a different scale of atomic interactions. In this way, both covalent interactions and non-covalent interactions are incorporated into the molecular representation on an equal footing. We systematically test Mol-GDL on fourteen commonly-used benchmark datasets. The results show that our Mol-GDL can achieve a better performance than state-of-the-art (SOTA) methods.

\section{Results}

\begin{figure*}
\centering
\includegraphics[width=0.9\textwidth]{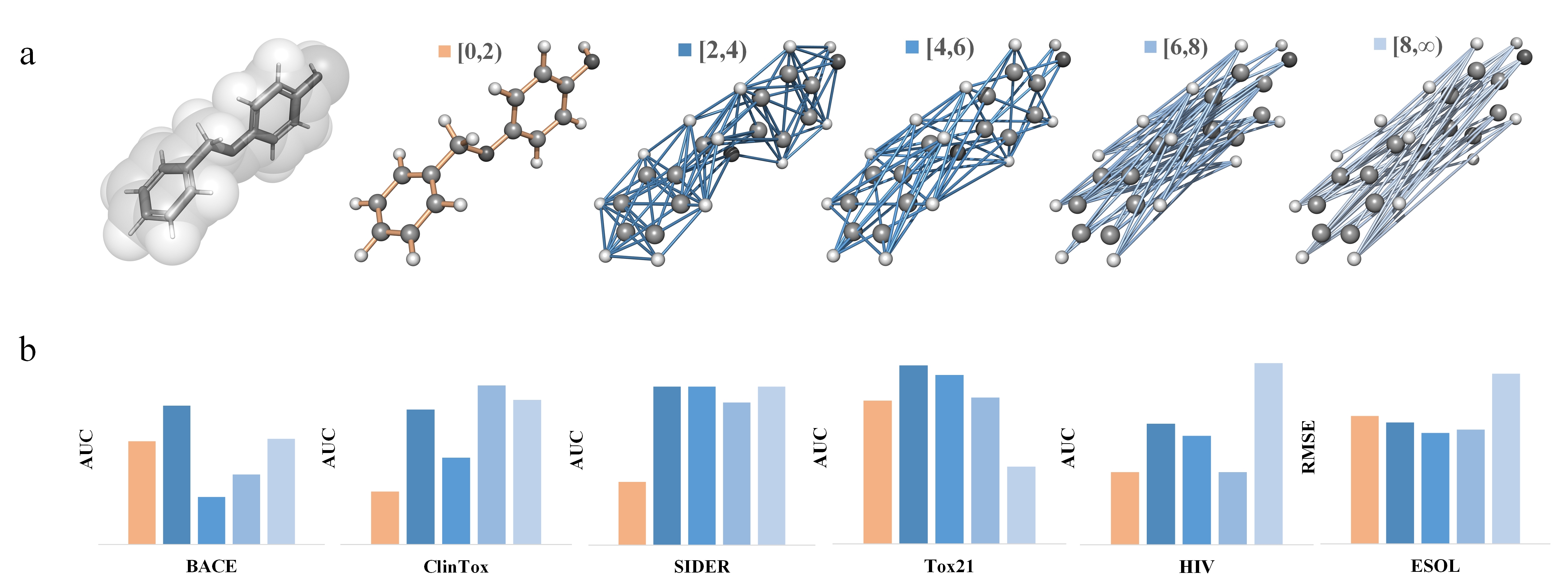}
\caption{
\textbf{Illustration of different molecular graph representations and the performance of their GDLs in six commonly-used datasets. The de facto standard of covalent-bond based molecular graph representation has clear limitations and is inferior to models with only non-covalent interactions.} \textbf{a} Molecular graph representations for Monobenzone molecule. The de facto standard of covalent-bond model is represented in orange color and the other four non-covalent-interaction-based graphs are in blue colors.  \textbf{b} The performance of GDLs with five different molecular representations on six most commonly-used datasets. The color of the bar (for each model) is the same as that of the corresponding molecular graph. For instance, the orange bars are for GDLs with covalent-bond-based molecular graph. } \label{fig:Noncovalent-GDL}
\end{figure*}

\subsection{Is covalent-bond based molecular graph a de facto standard for GDL?}\label{sec:molecular graph representation}
Currently, the de facto standard for molecular representation at the atomic level is the covalent-bond-based molecular graphs. Here we show that, in molecular property prediction, GDLs with molecular graphs constructed only from non-covalent interactions can achieve similar or even better results. We consider five different molecular graph representations and their GDL performance on the six most-commonly used datasets, including BACE, ClinTox, SIDER, Tox21, HIV, and ESOL, as illustrated in Figure \ref{fig:Noncovalent-GDL}. Of the five different molecular graphs, one is the de facto standard of the covalent-bond-based model and the other four are all constructed using only non-covalent interactions. Stated differently, all the edges in these four molecular graphs represent only non-covalent information and none of them are generated from the covalent bonds. Mathematically, these four non-covalent molecular graphs are constructed by defining the edges only between atoms within a certain pre-defined Euclidean distance (larger than the covalent bond distances). We specify a certain domain $I$ for each graph in a way that the edge in this graph exists if and only if the distance between the corresponding two atoms is in the domain $I$. For instance, a graph $G(I)$ with $I=[4\AA, 6\AA)$ means all the edges in graph $G(I)$ have lengths between 4\AA~ and 6\AA. Details for the molecular graph representation can be found in Section \ref{sec:Mol-GDL}.

It can be seen from Figure \ref{fig:Noncovalent-GDL} that GDL with the de facto standard of covalent-bond-based molecular graph does not have the best performance. In contrast, GDLs with non-covalent molecular graphs can not only have comparable results but even outperform the covalent-bond-based model. In fact, for all six datasets, the non-covalent model with $I=[4 \AA, 6\AA)$ has superior performance than the de facto standard model. More interestingly, even for the highly non-traditional molecular graph representation with $I=[8 \AA, \infty)$, in which edges only form between atoms that have Euclidean distances larger than 8 \AA, the corresponding GDL model can still have a comparably good performance and even outperform the de facto standard model in the four test datasets, including BACE, ClinTox, SIDER, and HIV. This demonstrates the great potential of novel molecular representations beyond the de facto standard of covalent-bond-based molecular graphs.

\subsection{Molecular geometric deep learning (Mol-GDL)} \label{sec:Mol-GDL}

\paragraph{Molecular graph representation for Mol-GDL}

For a molecule with $N$ atoms and atom coordinates denoted as ${\bf r}_1, {\bf r}_2, ..., {\bf r}_N$, its molecular graph representation can be expressed as $G(I)=(V, E(I))$. Here $I$ is a certain interaction region we used to define our graph representation. For simplicity, we only consider region in terms of $I=[x_{\rm min}, x_{\rm max})$ with $x_{\rm min}$ and $x_{\rm max}$ real values (that satisfy $0\leq x_{\rm min} <x_{\rm max}$). The set of nodes (or atoms) is denoted as $V$ and the set of edges is denoted $E(I)$. Note that each atom is modeled as a single node (or vertex) in our molecular graph representation. The adjacent matrix $A(I)= \left( a(I)_{ij} \right)_{ 1 \leq i \leq N; 1 \leq j \leq N}$ and its element $a(I)_{ij}$ is represented as,
\begin{equation}\label{eq:adj}
a(I)_{ij} =  \begin{cases}
1 , &  x_{\rm min} \leq  \|{\bf r}_i-{\bf r}_j\| <x_{\rm max} ~{\rm and} ~i \neq j\\
0, & \rm{others}.
\end{cases}
\end{equation}
Geometrically, it means that an edge is formed between the $i$-th and $j$-th atom if their Euclidean distance $\|{\bf r}_i-{\bf r}_j\|$ is within the domain of $I$, in molecular graph $G(I)$. For instance, we can set $I_0=[0\AA, 2\AA)$, the corresponding molecular graph $G(I_0)$ is exactly the de facto standard molecular graph, which is constructed with only covalent bonds, as their bond lengths are all within the region of $I_0=[0\AA, 2\AA)$. Mathematically, by the variation of the domain $I$, graph representations with dramatically different topologies can be generated. For instance, if we let $x_{\rm min}>2\AA$ (thus $x_{\rm max}>2\AA$), the corresponding molecular graph will contain only non-covalent interactions.

In our Mol-GDL, instead of using only one molecular graph, a series of graphs $G(I_k)$ are systematically generated by selecting different regions $I_k$ $(k>0)$ (See Method for details). Geometrically, for a certain graph $G(I_k)$, an edge is formed between two atoms only when their Euclidian distance is within the region of $I_k$. In this way, we have great flexibility to construct molecular graphs.

\paragraph{Geometric node features for Mol-GDL}

\begin{figure}[ht]
\centering
\includegraphics[width=0.55\textwidth]{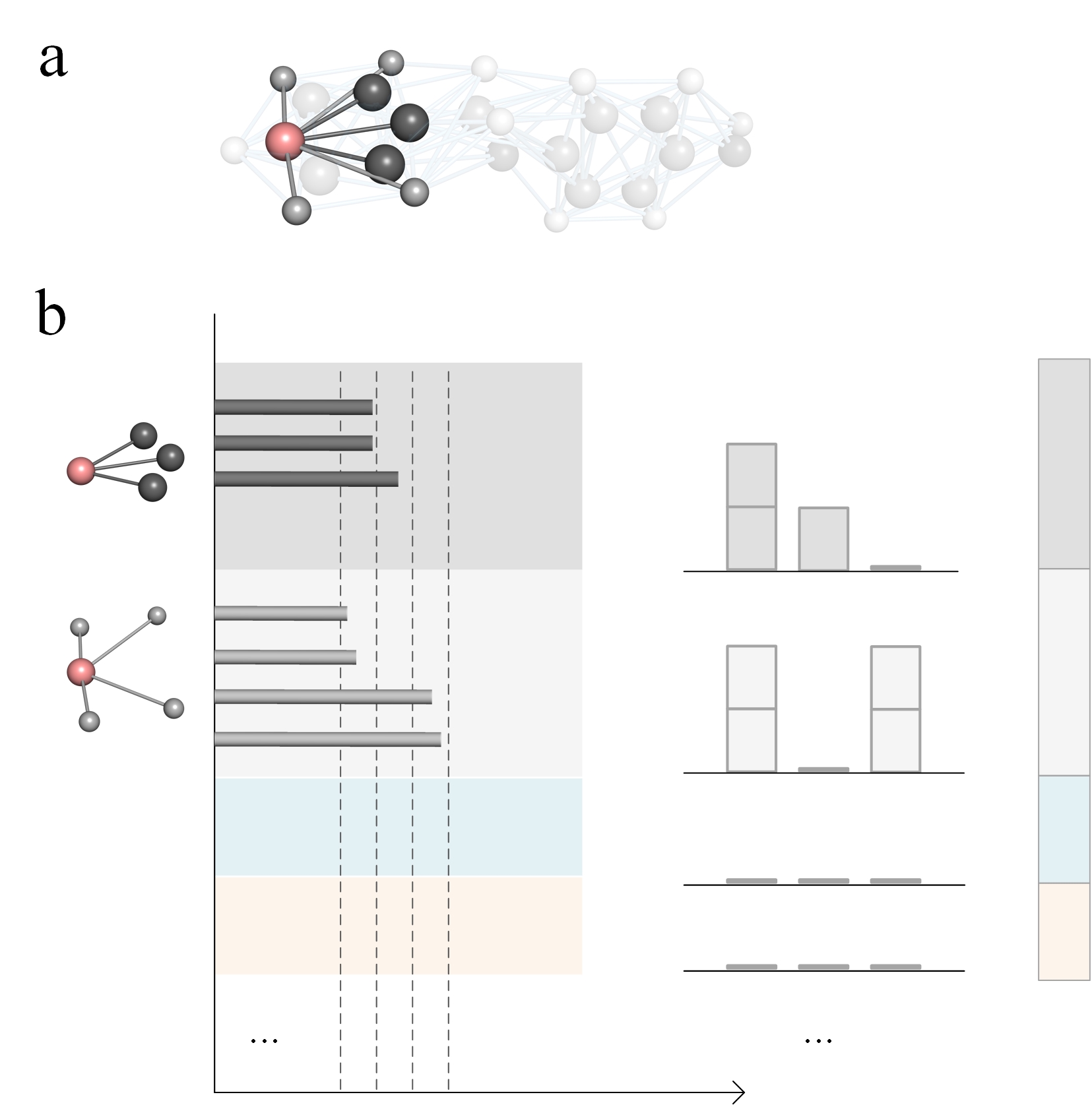}
\caption{\label{setup}
\textbf{Illustration of geometric node features for Mol-GDL. Different from all previous models, our geometric node features are solely determined by atom types and Euclidean distances between atoms.} \textbf{a} The illustration of a carbon atom (in pink) and its neighboring Carbon atoms (in dark black) and Hydrogen atoms (in light grey) from a molecular graph for Monobenzone. \textbf{b} In our geometric node features, the neighboring atoms are grouped based on their atom types. Here the two Carbon atoms are classified into one group and the four Hydrogen atoms are classified into the other. For each group, the Euclidian distances between all the neighboring atoms to the Carbon atoms are classified into several intervals. For each interval, we count the total number (or frequency) of distances within it. Here three equal-sized intervals are considered. For Carbon atoms, their frequencies in these intervals are (2, 1, 0), and for hydrogen atoms, their number is (2, 0, 2). These frequency numbers are then concatenated (in a predefined order according to atom types) into a fixed-length long vector, i.e., our geometric node feature.
}\label{fig:node feature}
\end{figure}
The other important setting for our Mol-GDL is the distance-related node features. In contrast with traditional node features, our node features contain only atom types and distance information.

Mathematically, the node feature for the $i$-th vertex of molecular graph $G(I)$ is denoted as $f_i(I)=[f_i(I, \alpha_1), f_i(I, \alpha_2),...,f_i(I, \alpha_m)]$. Note that here $\alpha_j$ (with $1\leq j\leq m$) means the type of atoms, such as carbon (C), nitrogen (N), oxygen (O), hydrogen (H), sulfur (S), etc. The element $f_i(I, \alpha_j)$ is defined as the number (or frequency) of edges in $G(I)$ that are formed between the $i$-th node and any other nodes that are of atom type $\alpha_j$. Mathematically, it is defined as follows,
            \begin{equation} \label{eq:Node_vector}
               f_i(I, \alpha_j) = \sum_{T_j=\alpha_j} \chi (x_{\rm min}\le ||{\bf r}_i-{\bf r}_j|| < x_{\rm max}),
            \end{equation}
here $T_j$ is the atom type of $j$-th atom and the value of the indicator function $\chi$ is 1 if the following condition is satisfied and 0 otherwise. Geometrically, the node feature contains $m$ descriptors and each descriptor represents the total number of edges connecting to atoms of a specific atom type. Figure \ref{fig:node feature} illustrates the node features using an carbon atom as an example.

Further, a refined node feature can be generated through the subdivision of domain $I$. More specifically, we can divide $I$ into $L$ intervals $\{ I^l=[x_{l-1}, x_l); l=1, 2,.., L\}$ with $x_0=x_{\rm min}$ and $x_L=x_{\rm max}$. The node feature element $f_i(I, \alpha_j)$ will be extended into a vector $[f_i(I^1, \alpha_j),f_i(I^2, \alpha_j),..., f_i(I^L, \alpha_j)]$. Similar, the scale element $f_i(I^1, \alpha_j)$ is defined as
  \begin{equation} \label{eq:Node_vector_v2}
               f_i(I^l, \alpha_j) = \sum_{T_j=\alpha_j} \chi (x_{l-1}\le ||{\bf r}_i-{\bf r}_j|| < x_l).
            \end{equation}
Note that our node features are solely determined by atom types and Euclidean distances between atoms.

\begin{figure*}
\centering
\includegraphics[width=0.9\textwidth]{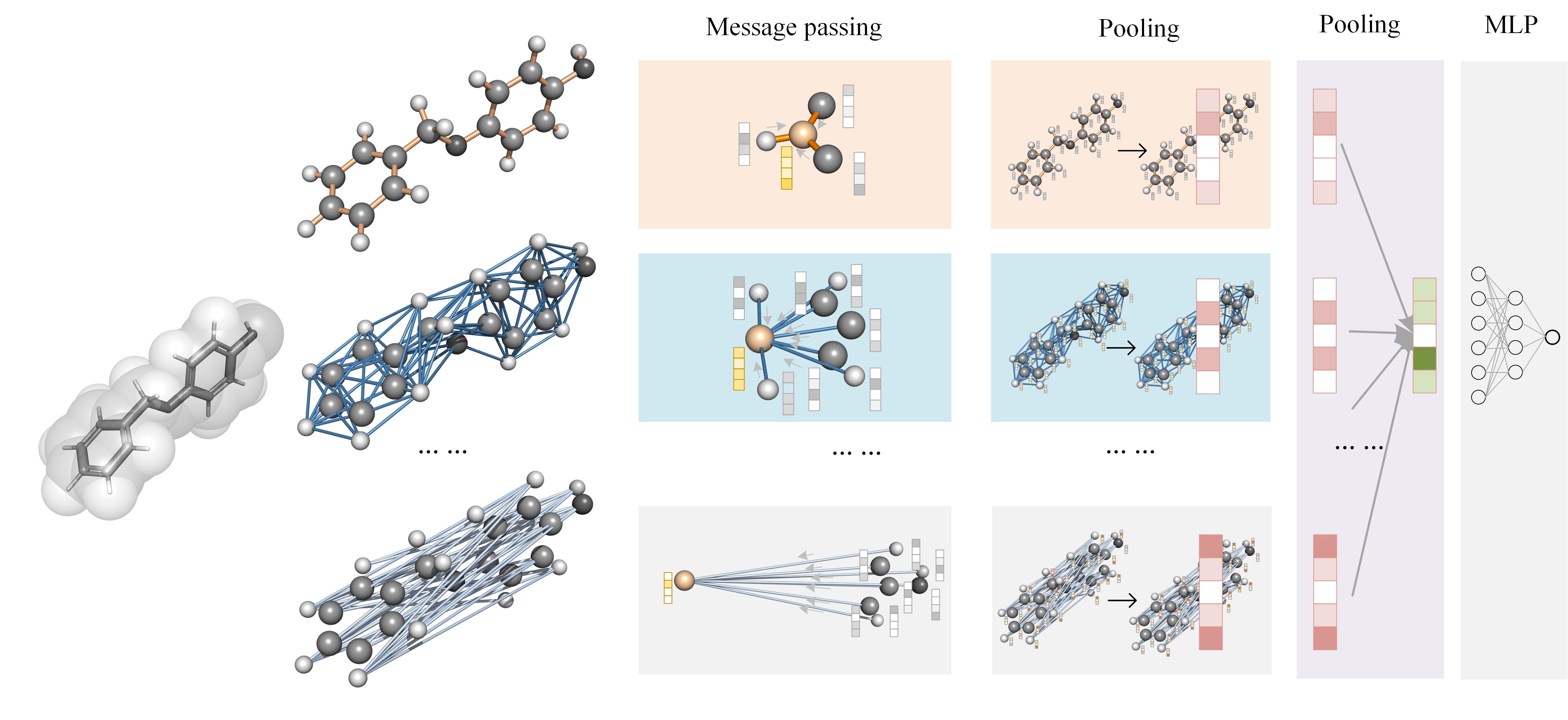}
\caption{\label{setup}
\textbf{The flowchart of our Mol-GDL model.} A set of molecular graphs are systematically constructed for each molecule. The geometric node features goes through the same message passing module. The two pooling operations, one is done within graphs and the other between groups, are used to aggregate individual node features into a single molecular feature vector. A multi-layer perceptron (MLP) is employed on the molecular feature vector to generate the final prediction. }\label{fig:overview of Mol-GDL}
\end{figure*}

\paragraph{The general framework for Mol-GDL}
Our Mol-GDL provides a geometric deep learning architecture that can comprehensively learn the multiscale information within a molecule. Figure \ref{fig:overview of Mol-GDL} illustrates the flowchart of our Mol-GDL. Different from all previous GDLs, a series of molecular graphs focusing on different scales of interactions are systematically constructed. A common message passing (MP) module is employed on each individual molecular graph. After that, two pooling operations are used. The first pooling is done at atomic level. For each molecular graph, node features are aggregated together into a single molecular feature vector. Second pooling is done at molecular graph level. Each molecular feature vector from the first pooling goes through a single-layer perceptron and then they are concatenated into a single feature vector. Finally, the single feature vector, which contains information from all the molecular graphs, goes through multi-layer perceptron (MLP) to generate the final prediction.

\subsection{Performance of Mol-GDL for molecular data analysis}
In this section, we present the comparison results between our Mol-GDLs and SOTAs on different types of molecular data analysis, including classification tasks on the molecular property, regression tasks on the molecular property, and tasks on molecular interaction prediction. The detailed information of datasets, SOTA models, and evaluation metrics can be found in Supplementary Note 1 and Table S1.

\paragraph{Mol-GDL for classification tasks on molecular property}

In the classification task, seven commonly used datasets\cite{wu2018moleculenet} are considered. These datasets can be roughly classified into two categories, one for biophysical properties (BACE, HIV, and MUV) and the other for physiological properties (BBBP, Tox21, SIDER, and ClinTox). More specifically, the BACE dataset provides quantitative binding results for inhibitors of human $\beta$-secretase 1(BACE-1)\cite{subramanian2016computational}. Its average number of atoms per molecule is around 65, which is the largest of the seven data sets. The HIV data set is for the study of the molecule's ability to inhibit HIV replication. The MUV data set, which is screened from PubChem BioAssay by applying a nearest neighbor analysis\cite{rohrer2009maximum}, is devoted to the validation of virtual screening techniques. The BBBP dataset is for barrier permeability\cite{martins2012bayesian}, while SIDER is a database of marketed drugs and adverse drug reactions\cite{kuhn2016sider}. Both Tox21 and ClinTox datasets\cite{gayvert2016data} are related to the toxicity of compounds and they all contain multiply classification tasks. The measurement of area under curve (AUC) is used for the evaluation of the results of the models. The detailed setting of our Mol-GDL parameters can be found in Supplementary Note 2.

The overall performance of Mol-GDL on classification benchmarks along with SOTAs is shown in Table \ref{tab:classification tasks}. It can be seen that our Mol-GDL has achieved the best results, and consistently outperform the SOTAs in all the tasks except only one from MUV. Our Mol-GDL on BACE is 0.863, which is better than the previous best result (AUC=0.856). The AUC results of the Mol-GDL model on the two datasets of HIV and MUV are 0.808 and 0.675, respectively. The results of Mol-GDL on BBBP and SIDER are significantly better than SOTAs. Among them, the performance on the SIDER dataset is the most prominent, and the AUC value of the Mol-GDL model is 20.68\% higher than the previous SOTA results. The performance of the Mol-GDL model on these two datasets of Tox21 and ClinTox is also much better than the comparison methods. It is worth noting that our model not only has the largest AUC value on the ClinTox dataset (AUC=0.966) but also has a much lower standard deviation (std=0.002) than all other existing methods.

\begin{table*}
\centering
\caption{\textbf{The comparison of Mol-GDL with SOTAs on seven commonly-used datasets, which contain only classification tasks on molecular properties. } Note that the subindex indicates standard deviation values. For instance, the element $65_{(19)}$ means the number of average atoms in BACE is 65 with 19 as its standard deviation. }\label{tab:classification tasks}
\resizebox{\linewidth}{!}{
\begin{tabular}{ llllllll }
\hline
Dataset & BACE & BBBP & ClinTox & SIDER & tox21 & HIV & MUV \\
No.molecule  & 1513	& 2039 & 1478 &	1427 &	7831 &	41127 &	93087 \\
No.average atoms & 65$_{(19)}$ & 46$_{(21)}$ & 50.58$_{(31)}$ & 65$_{(93)}$ & 36$_{(23)}$ & 46$_{(24)}$ & 43$_{(10)}$\\
No.tasks & 1 & 1 & 2 & 27 & 12 & 1 & 17\\
 \hline
D-MPNN\cite{yang2019analyzing} & 0.809$_{(0.006)}$ & 0.710$_{(0.003)}$ & 0.906$_{(0.006)}$ & 0.570$_{(0.007)}$ & 0.759$_{(0.007)}$ & 0.771$_{(0.005)}$ & 0.786$_{(0.014)}$ \\
AttentiveFP\cite{xiong2019pushing} & 0.784$_{(0.022)}$ & 0.643$_{(0.018)}$ & 0.847$_{(0.003)}$ & 0.606$_{(0.032)}$ & 0.761$_{(0.005)}$ & 0.757$_{(0.014)}$ & 0.766$_{(0.015)}$ \\
N-Gram$_{RF}$\cite{liu2019n} & 0.779$_{(0.015)}$ & 0.697$_{(0.006)}$ & 0.775$_{(0.040)}$ & 0.668$_{(0.007)}$ & 0.743$_{(0.004)}$ & 0.772$_{(0.001)}$ & 0.769$_{(0.007)}$ \\
N-Gram$_{XGB}$\cite{liu2019n} & 0.791$_{(0.013)}$ & 0.691$_{(0.008)}$ & 0.875$_{(0.027)}$ & 0.655$_{(0.007)}$ & 0.758$_{(0.009)}$ & 0.787$_{(0.004)}$ & 0.748$_{(0.002)}$ \\
PretrainGNN\cite{hu2019strategies} & 0.845$_{(0.007)}$ & 0.687$_{(0.013)}$ & 0.726$_{(0.015)}$ & 0.627$_{(0.008)}$ & 0.781$_{(0.006)}$ & 0.799$_{(0.007)}$ & 0.813$_{(0.021)}$ \\
GROVE$_{base}$\cite{rong2020self} & 0.826$_{(0.007)}$ & 0.700$_{(0.001)}$ & 0.812$_{(0.030)}$ & 0.648$_{(0.006)}$ & 0.743$_{(0.001)}$ & 0.625$_{(0.009)}$ & 0.673$_{(0.018)}$ \\
GROVE$_{large}$\cite{rong2020self} & 0.810$_{(0.014)}$ & 0.695$_{(0.001)}$ & 0.762$_{(0.037)}$ & 0.654$_{(0.001)}$ & 0.735$_{(0.001)}$ & 0.682$_{(0.011)}$ & 0.673$_{(0.018)}$ \\
GEM\cite{fang2022geometry} & 0.856$_{(0.011)}$ & 0.724$_{(0.004)}$ & 0.901$_{(0.013)}$ & 0.672$_{(0.004)}$ & 0.781$_{(0.001)}$ & 0.806$_{(0.009)}$ & \textbf{0.817}$_{(0.005)}$ \\
Mol-GDL & \textbf{0.863}$_{(0.019)}$ & \textbf{0.728}$_{(0.019)}$ & \textbf{0.966}$_{(0.002)}$ & \textbf{0.831}$_{(0.002)}$ & \textbf{0.794}$_{(0.005)}$ & \textbf{0.808}$_{(0.007)}$ & 0.675$_{(0.014)}$ \\

 \hline
\end{tabular}
}
\end{table*}

\paragraph{Mol-GDL for regression tasks on the molecular property}

The commonly used datasets for the regression task on the molecular property are mainly divided into two categories \cite{wu2018moleculenet}. One is for predicting the physical and chemical properties of molecules, including ESOL, FreeSolv, and Lipo. The other is in the field of quantum chemistry, including QM7, QM8, and QM9. The detailed setting of our Mol-GDL parameters can be found in Supplementary Note 2.

For a better comparison with SOTAs \cite{wu2018moleculenet, fang2022geometry}, we consider two metrics, root mean squre error (RMSE) and mean absolute error (MAE) respectively for the two types of regression tasks. Table \ref{tab:regression tasks} presents the results of the Mol-GDL model and SOTAs. Generally speaking, our Mol-GDL can achieve similar or even better results than SOTA. In particular, our Mol-GDL has the smallest RMSE values on the ESOL and FreeSolv tasks. On the QM7 dataset, the MAE value of the Mol-GDL model is 62.2, second only to GEM\cite{fang2022geometry}. Although the performance of Mol-GDL on the three datasets (Lipo, QM8, and QM9) did not rank among the top, it still beats several classical prediction methods of molecular properties, such as $N$-Gram$_{RF}$\cite{liu2019n}, $N$-Gram$_{XGB}$\cite{liu2019n}, GROVE$_{base}$\cite{rong2020self} and GROVE$_{large}$\cite{rong2020self}. The reason why Mol-GDL performs poorly on some regression tasks may be that the graph constructed by 3D coordinates is incomplete. As described in the previous literature\cite{pozdnyakov2022incompleteness,pozdnyakov2020incompleteness}, there are some structures in the 3D atomic cloud that cannot be distinguished by the distance-based GNN model, which leads to the suboptimal performance of Mol-GDL on these tasks.

\begin{table*}
\centering
\caption{\textbf{The comparison of Mol-GDL with SOTAs on six commonly-used datasets, which contain only region tasks on molecular properties.}Note that the subindex indicates standard deviation values.}\label{tab:regression tasks}
\resizebox{\linewidth}{!}{
\begin{tabular}{ llllllll }
\hline
&\multicolumn{3}{l}{RMSE}&\multicolumn{3}{l}{MAE}\\
\cline{2-7}
Dataset & ESOL & FreeSolv & Lipo & QM7 & QM8 & QM9 \\
No.molecules & 1128 & 642 & 4200 & 6830 & 21786 & 133885\\
No.average atoms & 26$_{(13)}$ & 18$_{(7)}$ & 49$_{(15)}$ & 16$_{(3)}$ & 	16$_{(3)}$ & 18$_{(3)}$ \\
No.tasks & 1 & 1 & 1 & 1 & 12 & 12\\
 \hline
D-MPNN\cite{yang2019analyzing} & 1.050$_{(0.008)}$ & 2.082$_{(0.082)}$ & 0.683$_{(0.016)}$ & 103.5$_{(8.6)}$ & 0.0190$_{(0.0001)}$ & 0.00814$_{(0.00001)}$\\
AttentiveFP\cite{xiong2019pushing} & 0.877$_{(0.029)}$ & 2.073$_{(0.183)}$ & 0.721$_{(0.001)}$ & 72.0$_{(2.7)}$ & 0.0179$_{(0.0001)}$ & 0.00812$_{(0.00001)}$\\
N-Gram$_{RF}$\cite{liu2019n} & 1.074$_{(0.107)}$ & 2.688$_{(0.085)}$ & 0.812$_{(0.028)}$ & 92.8$_{(4.0)}$ & 0.0236$_{(0.0006)}$ & 0.01037$_{(0.00016)}$\\
N-Gram$_{XGB}$\cite{liu2019n} & 1.083$_{(0.082)}$ & 5.061$_{(0.744)}$ & 2.072$_{(0.030)}$ & 81.9$_{(1.9)}$ & 0.0215$_{(0.0005)}$ & 0.00964$_{(0.00031)}$\\
PretrainGNN\cite{hu2019strategies} & 1.100$_{(0.006)}$ & 2.764$_{(0.002)}$ & 0.739$_{(0.003)}$ & 113.2$_{(0.6)}$ & 0.0200$_{(0.0001)}$ & 0.00922$_{(0.00004)}$\\
GROVE$_{base}$\cite{rong2020self} & 0.983$_{(0.090)}$ & 2.176$_{(0.052)}$ & 0.817$_{(0.008)}$ & 94.5$_{(3.8)}$ & 0.0218$_{(0.0004)}$ & 0.00984$_{(0.00055)}$\\
GROVE$_{large}$\cite{rong2020self} & 0.895$_{(0.017)}$ & 2.272$_{(0.051)}$ & 0.823$_{(0.010)}$ & 92.0$_{(0.9)}$ & 0.0224$_{(0.0003)}$ & 0.00986$_{(0.00025)}$\\
GEM\cite{fang2022geometry} & 0.798$_{(0.029)}$ & 1.877$_{(0.094)}$ & \textbf{0.660}$_{(0.008)}$ & \textbf{58.9}$_{(0.8)}$ & \textbf{0.0171}$_{(0.0001)}$ & \textbf{0.00746}$_{(0.00001)}$\\
Mol-GDL & \textbf{0.798}$_{(0.024)}$ & \textbf{1.809}$_{(0.100)}$ & 0.779$_{(0.007)}$ & 62.2$_{(0.4)}$ & 0.0205$_{(0.0001)}$ & 0.00952$_{(0.00013)}$\\

 \hline
\end{tabular}
}
\end{table*}

\paragraph{Mol-GDL for molecular interaction analysis}

In this task, we use the dataset from DeepDDS\cite{wang2022deepdds}, which contains 36 anticancer drugs, 31 human cancer cell lines, and 12,415 unique drug pair-cell line combinations. Six different validation metrics are used to measure the performance of models, including AUC, AUPR, Recall, Precision and F1-score. Note that a slightly different architecture and parameter setting is used and the details cane be found in Supplementary Note 3 and Table S4.

Five deep learning-based drug synergy prediction methods, including TranSynergy\cite{liu2021transynergy}, DeepSynergy\cite{preuer2018deepsynergy}, Deep Tensor Factorization(DTF)\cite{sun2020dtf}, DeepDDS$_{GAT}$\cite{wang2022deepdds}, and DeepDDS$_{GCN}$\cite{wang2022deepdds}, and two machine learning methods, i.e., XGBoost\cite{chen2016xgboost} and Random Forest (RF)\cite{ho1995random}, are used as baselines. Since the task in the dataset is drug combination prediction between two drugs, a slightly different learning architecture is considered in our Mol-GDL.  Table \ref{tab:drug combination} shows the overall performance of the Mol-GDL model in predicting synergistic drug combinations. It can be seen that our Mol-GDL model outperforms all the other methods on five validation metrics.


\begin{table}
\centering
\caption{\textbf{The comparison of the performance of Mol-GDL and SOTAs on synergistic drug combination dataset.} Note that the subindex indicates standard deviation values.}\label{tab:drug combination}
\begin{tabular}{llllll}
\hline
\multirow{2}{*}{Methods} & \multicolumn{4}{l}{Metrics}\\
\cline{2-6}
&AUC & AUPR & REC & PREC & F1\\
 \hline
XGBoost\cite{chen2016xgboost} & 0.92$_{(0.01)}$ & 0.92$_{(0.01)}$ & 0.84$_{(0.01)}$ & 0.84$_{(0.01)}$ & 0.84$_{(0.01)}$\\
RF\cite{ho1995random} & 0.86$_{(0.01)}$ & 0.85$_{(0.02)}$ & 0.74$_{(0.01)}$ & 0.78$_{(0.02)}$ & 0.76$_{(0.01)}$\\
TranSynergy\cite{liu2021transynergy} & 0.90$_{(0.01)}$ & 0.89$_{(0.01)}$ & 0.80$_{(0.01)}$ & 0.84$_{(0.01)}$ & 0.82$_{(0.01)}$\\
DTF\cite{sun2020dtf} & 0.89$_{(0.01)}$ & 0.88$_{(0.01)}$ & 0.77$_{(0.03)}$ & 0.82$_{(0.01)}$ & 0.80$_{(0.02)}$\\
DeepSynergy\cite{preuer2018deepsynergy} & 0.88$_{(0.01)}$ & 0.87$_{(0.01)}$ & 0.75$_{(0.01)}$ & 0.81$_{(0.01)}$ & 0.78$_{(0.01)}$\\
DeepDDS$_{GAT}$\cite{wang2022deepdds} & 0.93$_{(0.01)}$ & 0.93$_{(0.01)}$ & 0.84$_{(0.07)}$ & 0.85$_{(0.07)}$ & 0.85$_{(0.07)}$\\
DeepDDS$_{GCN}$\cite{wang2022deepdds} & 0.93$_{(0.01)}$ & 0.92$_{(0.01)}$ & 0.84$_{(0.01)}$ & 0.85$_{(0.01)}$ & 0.84$_{(0.01)}$\\
Mol-GDL & \textbf{0.94}$_{(0.01)}$ & \textbf{0.94}$_{(0.01)}$ & \textbf{0.86}$_{(0.01)}$ & \textbf{0.86}$_{(0.01)}$ & \textbf{0.86}$_{(0.01)}$\\
 \hline
\end{tabular}
\end{table}

\section{Discussion}

\subsection{Incorporation of non-covalent-bond information enhances performance}
The role of non-covalent bonds in predicting molecular properties has long been underappreciated. In this work, we systematically compare the performance of GDLs from non-covalent-bond-based molecular graphs and covalent-bond-based molecular graphs in 13 datasets. The detailed comparison setting is discussed in Supplementary Note 4 and the results are listed in Table S5. It can be seen that the performance of our Mol-GDLs is better than that of the models that consider only covalent-bond information on most datasets, indicating that non-covalent bonds provide a positive impact on improving the prediction performance of the model. Therefore, these results show the importance of non-covalent interactions in molecular property prediction.

Further, we consider some extreme cases to explore the effects of covalent and non-covalent bonds in molecular representations for GDLs. Three types of neighboring relations are used for the construction of molecular graphs, i.e., $n$-nearest neighbor, $n$-farthest neighbor, and $n$-random neighbor. Here $n$ is an integer number and can be taken from 1 to 7. The results are shown in Tables S6 and S7, and we have the following observations. First, as the number of neighbor nodes increases, the performance will get better and then tend to be stable. Second, with only a few farthest neighbors or random neighbors, i.e., $n$=1 or 2, the model can still have comparably good prediction, in particular the classification tasks. Third, on some datasets, the prediction performance of molecular graphs constructed by the $n$ farthest neighbors is better than that of the $n$ nearest neighbors, in particular for BBBP, ClinTox, and QM7 datasets. These observations again demonstrate the importance of non-covalent interactions and the potential novel molecular representations beyond the de facto standard of covalent-bond-based molecular graphs.

\subsection{Geometric node features enhance GDL performance}
\begin{figure}
\centering
\includegraphics[width=0.70\textwidth]{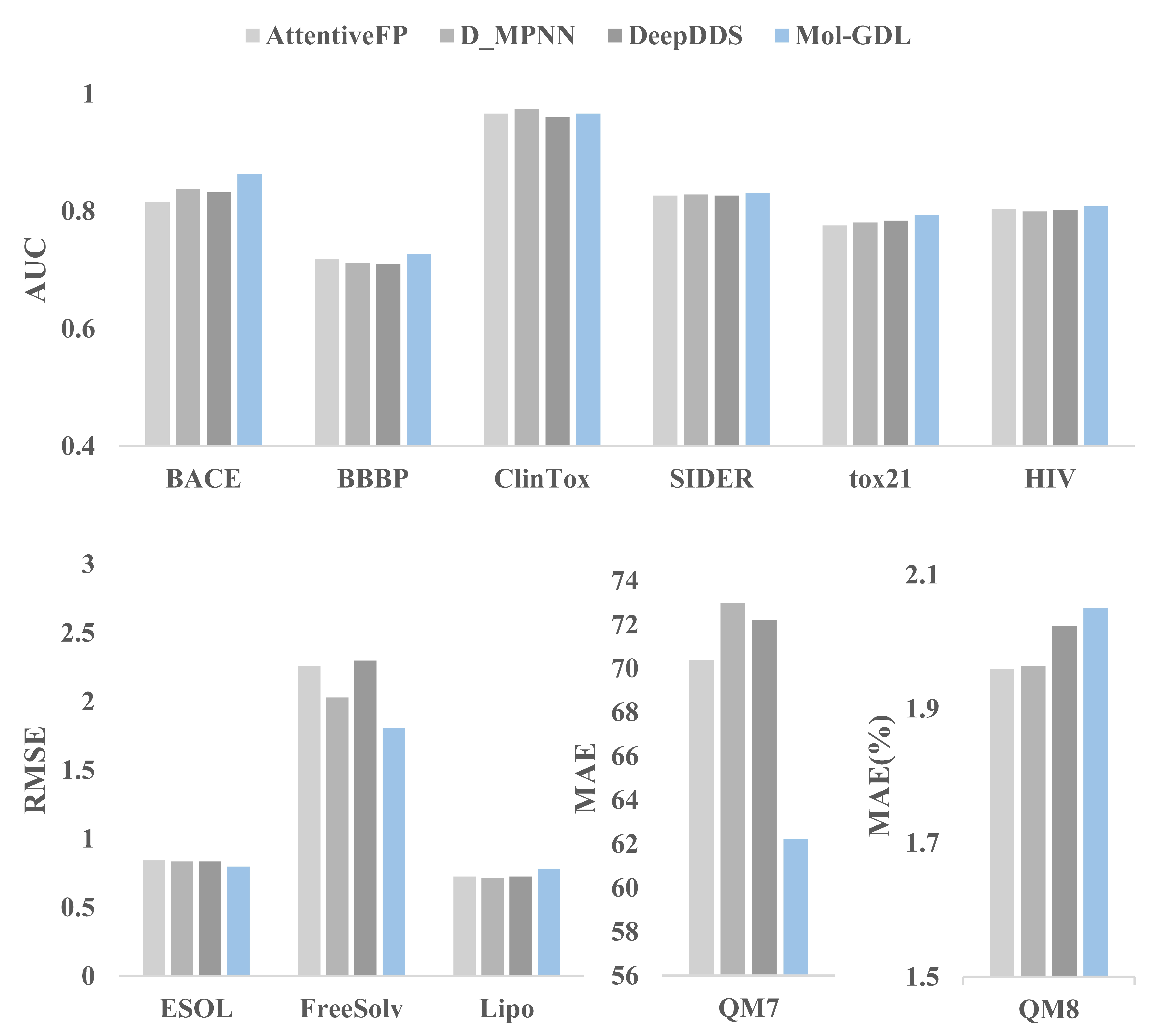}
\caption{\label{fig:Figure3}
\textbf{Comparison of different methods of calculating node feature.}}\label{fig:result of node feature}
\end{figure}

Note features are of key importance for GDLs. Here we propose the geometric node features that are related only to atomic types and Euclidean distances. To verify the efficiency of our geometric node feature approach, we compare it with three different feature engineering methods from the learning models of AttentiveFP\cite{xiong2019pushing}, D-MPNN\cite{yang2019analyzing,stokes2020deep} and DeepDDS\cite{wang2022deepdds}. Note that the features used in these three models are derived from the structural, physical, chemical, and biological properties, such as the number of atoms, normal charge, chirality, hybridization, aromaticity, etc. For a fair comparison, the same graph neural network architecture from our Mol-GDL is used. That is the same Mol-GDL architecture equipped with four types of different initial node features on 11 datasets (MUV and QM9 datasets are omitted due to memory issues). The detailed setting can be found in Supplementary Note 5 and the results are listed in Figure \ref{fig:result of node feature} and Table S8. It can be seen that our geometric node features are superior to the three feature engineering approaches in almost all the datasets. In particular, our geometric node feature approach has obvious advantages in BACE, BBBP, freesolv, and QM7 datasets.

It should be noted that our geometric node feature approach is developed based on strong relations between molecular structures and their functions. Even though our approach contains only the atomic types and Euclidean distances, various physical, chemical, and biological information has been implicitly incorporated into it. In fact, the positions of all the atoms within a molecule are the direct reflection of the overall interactions within the molecule. Stated differently, two atoms are close to each other only when the overall interactions (from all the physical, chemical, and biological effects) between them are relatively strong. Since this interaction information is implicitly embodied into its 3D structure through the structure-function relationships, an efficient molecular representation can be obtained as long as we fully explore the structural information of the molecule. From another point of view, the length of chemical bonds is used as the node feature in our Mol-GDL model. This is essentially the process of storing the information of non-covalent bonds in the initial representation. The great performance of our approach again indicates the importance of non-covalent interactions in molecular representation.

\subsection{General properties of Mol-GDL}
With the highly efficient molecular representation, our Mol-GDL model achieves a better performance than SOTAs on most datasets. The performance of our Mol-GDL model is highly consistent and does not depend on the size of the datasets. In Tables \ref{tab:classification tasks} and \ref{tab:regression tasks}, our Mol-GDL achieves the best performance in 8 of the 13 datasets. Especially on the classification task, the average performance is improved by about 3\% compared to the best model GEM in the SOTA methods, and 6 of the 7 classification datasets are optimal. Note that the number of samples varies widely in datasets for these classification tasks. There are only about 1,400 molecules in ClinTox and SIDER datasets, while more than 40,000 molecules are included in the HIV dataset.  Even with the huge size difference in the datasets, the performance of our Mol-GDL model is consistently better than SOTAs.

Further, the performance of our Mol-GDL model can be related to the size of the molecules. In particular, Mol-GDL may have slightly inferior results than SOTAs for small-sized molecules. In the regression task, although Mol-GDL model has better results than most of the models, such as N-Gram$_{RF}$, N-Gram$_{XGB}$, GROVE$_{base}$ and GROVE$_{large}$, it has slightly inferior results than GEM on QM7, QM8, and QM9. The average number of atoms per molecule for these three datasets is about 16, 16, and 18, respectively. However, it should be noticed that GEM is a pre-trained self-supervised model and it uses a much larger dataset for the pre-training process. In contrast, our Mol-GDL uses only the regular training set, which is significantly smaller in size than the dataset for pre-training. It is worth mentioning that the other three models, i.e., PretainGNN, GROVER$_{base}$, and GROVER$_{large}$, also use the pre-training process, but their performance is still inferior to our Mol-GDL for all the tasks.

Moreover, other than molecular property analysis, our Mol-GDL model can also achieve great performance in molecular interaction prediction. As shown in Table \ref{tab:drug combination}, our Mol-GDL outperforms SOTA methods for predicting synergistic drug combinations on all validation metrics. This indicates that efficient molecular graph representations are important to the analysis of not only properties at the single molecular level but also interactions between two or more molecules.

Finally, our Mol-GDL models can be further generalized or extended from several different aspects. First, we can use our Mol-GDL framework to generate a series of molecular fingerprints \cite{merkwirth2005automatic,duvenaud2015convolutional,coley2017convolutional,xu2017deep,winter2019learning}. More specifically, a series of molecular descriptors can be obtained from the aggregation of node features. Second, Mol-GDL based multi-task learning models can be constructed to further improve the performance of Mol-GDL. Third, distance GNN models have some intrinsic limitations to distinguish some special molecular structures \cite{pozdnyakov2022incompleteness}. Many-body descriptors alone are also incomplete for molecular structure representation \cite{pozdnyakov2020incompleteness}. The generalization of Mol-GDL into higher-order topological models, such as simplicial complex, polyhedron complex, hypergraph, etc, may provide ways for a more efficient and complete molecular representation.

\section{Conclusion}
Molecular representation learning plays an important role in molecular property prediction. Existing molecular property prediction work often relies on the de facto standard of covalent-bond-based molecular graphs for representing molecular topology at the atomic level, and totally ignores the non-covalent interactions within the molecule. In this study, we propose a molecular geometric deep learning(Mol-GDL) model to predict the properties of molecules, which aims to comprehensively consider the information of covalent and non-covalent interactions of molecules. Extensive tests have demonstrated the effectiveness of the Mol-GDL model, and the important role of non-covalent interactions in molecular property prediction has also been validated through multiple tests. In future work, we consider multi-task learning or incorporating more effective information other than non-covalent bonds to further improve the performance of Mol-GDL.

\section{Methods}

\subsection{Molecular graph representation for Mol-GDL}
Graph neural networks can encode molecular structural information into a low-dimensional representation vector from a high-dimensional space. It is a popular method to regard the atoms in the molecule as nodes and the chemical bonds as edges in the graph, and then use the method of message passing to predict the properties of the molecule\cite{stokes2020deep}\cite{li2020monn}. However, ignoring the quantitative distances between atoms in traditional graphs may result in the loss of a great deal of critical physical and chemical information in molecules. Molecular graphs reconstructed from specific distance intervals can capture some different physicochemical and biophysical patterns, such as hydrogen bonding and van der Waals interactions between different atoms\cite{meng2021persistent}\cite{chen2021algebraic}.

A molecule can be represented as a graph $G(I)=(V, E(I))$, where $E(I)$ is an edge set and $V$ is a node set. $I$ is a certain interaction region. 
Physically, the covalent bonds are usually within the interaction region of $[0\AA, 2\AA]$. Non-covalent interactions are comparably weak and are within a much larger range. For instance, the distance between donor and acceptor atoms in hydrogen bonds, is roughly between $[2.0\AA, 4.0\AA]$. Van der Waals forces can cover a much larger range and play an very important role in the range $[4.0\AA, 6.0\AA]$. The electrostatic forces can go further to $[6.0\AA, 8\AA]$. To fully explore the physicochemical and biophysical patterns in non-covalent interactions, we divide the interval region greater than 2$\AA$ into several non-overlapping regions. For each interaction region, a corresponding molecular graph is generated and its associated adjacency matrix is defined as Eq. (\ref{eq:adj}).

To balance the computational costs and model accuracy, a total of 4 segments are considered for non-covalent interactions, namely $I_{1} = [2\AA, 4\AA),~I_{2} = [4\AA, 6\AA),~I_{3} = [6\AA, 8\AA),~I_{4} = [8\AA,\infty)$. Note that the selection of interaction regions to build molecular graphs is not unique. For convenience, we let $I_{0} = [0\AA, 2\AA)$ denotes the interval of covalent interactions. In this way, total 5 interaction regions are considered and a total of 5 molecular graphs, denoted as $G=\left\{G_1,G_2,G_3,G_4,G_5\right\}$, are generated for each molecular structure, in all of our three types of tasks except those from QM7, QM8, and QM9. Since the number of atoms in the three datasets QM7, QM8, and QM9 is relatively small, the number of non-covalent interactions greater than $6\AA$ is very small. Therefore, we divide the non-covalent interaction interval for these three datasets into three segments, namely $I_{1} = [2\AA,4\AA),~I_{2} = [4\AA,6\AA),~I_{3} = [6\AA,\infty)$. Similarly, let $I_{0} = [0\AA,2\AA)$ represent the interval of covalent bonds. In this way, total 4 interaction regions are considered and a total of 4 molecular graphs, denoted as $G=\left\{G_1,G_2,G_3,G_4\right\}$,  are generated for each molecular structure from QM7, QM8, and QM9.



\subsection{Graph neural network architecture in Mol-GDL}
For a molecular $G(I_k)$, denoted as $G_k$ for short, we use message passing to learn the feature representation of each node,
\begin{align}\label{eq4}
\bm{h}_{i}^{G_{k},(t)} = \sigma\left( {\sum\limits_{j \in \mathcal{N}{(i)} \cup {\{ i\}}}{\frac{1}{\sqrt{d(i)} \cdot \sqrt{d(j)}} \cdot \left( {W_{G_{k}}\cdot \bm{h}_{j}^{G_{k},{({t-1})}}} \right)}} \right).
\end{align}

In the $t$-th iteration, the $i$-th node feature vector $h_i^{G_{k},(t)}$ of a graph $G_{k}$ is obtained by gathering node feature vectors of its neighbors, denoted as $\mathcal{N}{(i)}$, and itself, from $(t-1)$-th iteration. Here $d(\cdot)$ represents the node degree and $W_{G_{k}}$ is the weight matrix (to be learned). Computationally, we usually repeat the process 1 to 3 times, and the final node feature is denoted as $\bm{h}_{i}^{G_{k}}$.

After the message passing, we aggregate all node features in a molecular graph through a pooling process,
\begin{align}\label{eq5}
\bm{h}_{G_{k}} = Pooling\left(\bm{h}_{i}^{G_{k}} \middle| i \in [1, 2,..., N] \right)
\end{align}
where $Pooling(\cdot)$ is a pooling function, such as max pooling, mean pooling, etc. Note that this pooling process is within the single molecular graph.

In our Mol-GDL, each molecule will have more than one molecular graph, we aggregate all the features from its molecular graphs,
\begin{align}\label{eq6}
\bm{h}_{G_{k}}^{'} = \sigma\left( W_{G_{k}}^{'}\bm{h}_{G_{k}} + b_{G_{k}}^{'} \right)
\end{align}
\begin{align}\label{eq7}
\bm{h}_{G} = READOUT\left(\bm{h}_{G_{k}}^{'} \middle| k \in \left\{ {1,2,...,5} \right\} \right)
\end{align}
where $READOUT(\cdot)$ denotes a pooling function, and we choose concatenation (or mean) in this study.

Finally, a multiply layer perceptron (MLP) is utilized for the final prediction,
\begin{align}\label{eq8}
\hat{y} = \delta\left( W^{1}\sigma\left( {W^{0}\bm{h}_{G} + b^{0}} \right) + b^{1} \right)
\end{align}
where $\sigma\left(\cdot\right)$ and $\delta(\cdot)$ are $Relu\left(\cdot\right)$ and $Sigmoid\left(\cdot\right)$, respectively. For regression tasks, $\delta(\cdot)$ is a linear activation function. The $\ell_1$ loss and cross-entropy loss are implemented for regression and classification tasks, respectively.\\

\subsection{Geometric node features for Mol-GDL}
The geometric node features are of great importance for GDLs. In our Mol-GDL geometric node features that contain only atomic types and Euclidean distance information. Computationally, we consider a total of 12 types of atoms, including $C$, $H$, $O$, $N$, $P$, $Cl$, $F$, $Br$, $S$, $S_i$, $I$ and all the rest atoms as one type. These atoms are chosen due to their high frequencies in the molecules in our datasets. For the $i$-th atom, atom type $\alpha_j$ will contribute a component $f_i(I, \alpha_j)$ in the geometric node feature $f_i(I) = [f_i(I, \alpha_1), f_i(I, \alpha_2), ..., f_i(I, \alpha_12)]$. Here $f_i(I, \alpha_j)$ means the number (or frequency) of all the neighboring atoms of type $\alpha_j$ for the $i$-th atom. Further, different subdivision of interaction region $I$ is done for different types of atoms. More specifically, for $C$ (and $H$), its node feature vector components are of the size 19 after subdivision. That is the component $f_i(I, C)$ has been extended to a vector $[f_i(I^1, C), f_i(I^2, C),...,f_i(I^{19}, C)]$. Here $f_i(I^1, C)$ is the number of all the neighboring atoms of type $\alpha_j$ within the interaction subregion $I^1$ for the $i$-th atom. For $O$ and $N$, their node feature components are 4 after subdivision. All the rest types of atoms have the same node feature components of size 2 after subdivision. The detailed information for the subdivision of the interaction region (based on different atom types) is listed in Supplementary Note 6 and Table S9.


\section{Funding}
This work was supported in part by the Natural Science Foundation of China (NSFC grant no. 61873089, 62032007), Nanyang Technological University Startup Grant (grant no. M4081842), Singapore Ministry of Education Academic Research fund (grant no. Tier 1 RG109/19, MOE-T2EP20120-0013, MOE-T2EP20220-0010) and China Scholarship Council (CSC grant no. 202006130147).

\bibliographystyle{IEEEtran}
\bibliography{Mol-GDL_v9}

\end{document}